\documentclass[sn-mathphys,Numbered]{sn-jnl}

\usepackage{graphicx}%
\usepackage{multirow}%
\usepackage{amsmath,amssymb,amsfonts}%
\usepackage{amsthm}%
\usepackage{mathrsfs}%
\usepackage[title]{appendix}%
\usepackage{xcolor}%
\usepackage{textcomp}%
\usepackage{manyfoot}%
\usepackage{booktabs}%
\usepackage{algorithm}%
\usepackage{algorithmicx}%
\usepackage{algpseudocode}%
\usepackage{listings}%
\usepackage{bm}
\usepackage{soul}

\def\3he{$^3$He}
\def\4he{$^4$He}

\def\Aph{\textit{A} phase}
\def\Bph{\textit{B} phase}
\def\etal{\textit{et al.}}
\def\pwave{\textit{p}-wave}

\begin{document}

\title{Planar Aerogel and Superfluid \3he{}, Structure and Transitions}

\author{J. W. Scott}
\author{M.D. Nguyen}
\author{D. Park}
\author{W. P. Halperin}

\affil{\orgdiv{Department of Physics and Astronomy}, \orgname{Northwestern University}, \orgaddress{\street{2145 Sheridan Road}, \city{Evanston}, \postcode{60208}, \state{Illinois}, \country{USA}}}

\date{\today}

\abstract{
Anisotropic aerogel possesses structure which exhibits a strong influence over the composition and orientation of the order parameter of imbibed superfluid \3he{}. Computational studies have identified stretched aerogel with plane-like structures and compressed aerogel with nematic-like structures. Studies of the \Bph{} of superfluid \3he{} in stretched aerogel display an enhanced nuclear magnetic susceptibility likely caused by Andreev bound states near plane-like impurity sites. We report further details on the influence of these planar structures on both magnetic and orbital orientation transitions. The orbital orientation transitions appear in both the {\it B} and \Aph{}s of stretched and compressed aerogels. These transitions result from a crossover of the superfluid coherence length with long and short length scale structure with the coherence length and are consequently magnetic field independent. Additionally, the apparent temperature-independence of the susceptibility of the \Bph{} equal to that of the \Aph{} in stretched aerogel, is in marked contrast with the field dependence of the superfluid \textit{A} to \textit{B} phase transition, indicating that it is a near-isentropic transition.}


\maketitle

\section{Introduction}

As a \pwave{} superfluid, \3he{} is subject to pairbreaking via scattering from surfaces and impurity sites\cite{Lar.65,Thu.98}. As a consequence, the orbital component of the \3he{} superfluid order parameter couples strongly to anisotropic scattering\cite{Vol.13}. Dilute impurities in the form of aerogel, grown with a structural anisotropy, orient and alter the order parameter. Anisotropy has been previously shown to strongly influence the relative stability of the superfluid phases, with compressed aerogel favoring the \Bph{}\cite{Li.14b} with a nematic-like structure, and stretched aerogel favoring the \Aph{}\cite{Pol.12}  with a planar-like structure\cite{Ngu.24}. The more extreme case of nematic Nafen aerogel has likewise been shown to stabilize the polar phase\cite{Dmi.15a}. When imbibed in anisotropic silica aerogel an orbital flop transition has been observed\cite{Zim.18}, where the axis of the orbital angular momentum $\bm{\ell}$ spontaneously reorients.

\section{Experimental Details}

The stretched aerogel is grown intrinsically strained, with the strain characterized by the ratio of the diameter of the sample after supercritical methanol extraction, compared to the inner diameter of the tube in which it was grown\cite{Pol.08}.  The strain axis of the aerogel sample, $\bm{\epsilon}$ was measured via optical birefringence, and found to coincide with the cylindrical axis. The $16\%$ stretched aerogel sample is situated in a Stycast epoxy fixture attached to the silver heat exchanger by a fill line, and is surrounded by a quantity of pure, bulk \3he{}.

The identity and orientation of the superfluid phase of \3he{} can be established from the NMR resonance frequency and its dependence on tipping-angle caused by the nuclear dipole-dipole interaction\cite{Leg.73}, providing information on the structure and orientation of the superfluid order parameter\cite{Zim.19}. Additionally, the total spectral weight of the NMR spectrum is proportional to the susceptibility, measured as a function of temperature.  In these experiments, temperature sweeps were performed by delivering intermittent NMR pulses as the sample warmed. The temperature was monitored with a \3he{} melting curve thermometer and $^{195}\mathrm{Pt}$ NMR thermometry.

The measurements of spectra and susceptibility for $16\%$ stretched aerogel in Fig.\,\ref{fig:4panel}\,\textbf{a},\,\textbf{c}  were performed for a range of pressures and fields from $74.5\,\mathrm{mT}$ to $195\,\mathrm{mT}$ and from $19.0$ to $26.55\,\mathrm{bar}$\cite{Sco.23}. The sample gas had a small amount of \4he{} added such that the surface of the aerogel was coated with $\approx3.5$ atomic layers (commonly referred to as pre-plating), suppressing the formation of paramagnetic solid \3he{}\cite{Spr.96,Col.09}.  The sample was oriented such that $\bm{\epsilon}$ was aligned with the static magnetic field $\bm{H_{0}}$. We compare with measurements from a $19\%$ compressed aerogel Fig.\,\ref{fig:4panel}\,\textbf{b},\,\textbf{d}, performed and analyzed in the same way\cite{Zim.18}. It is significant that this aerogel has the opposite direction of strain and consequently a nematic structure as compared with the planar structure of the $16\%$ stretched sample.

\section{NMR in Anisotropic Aerogel}

\begin{figure*}
\centering
\includegraphics[width=120mm]{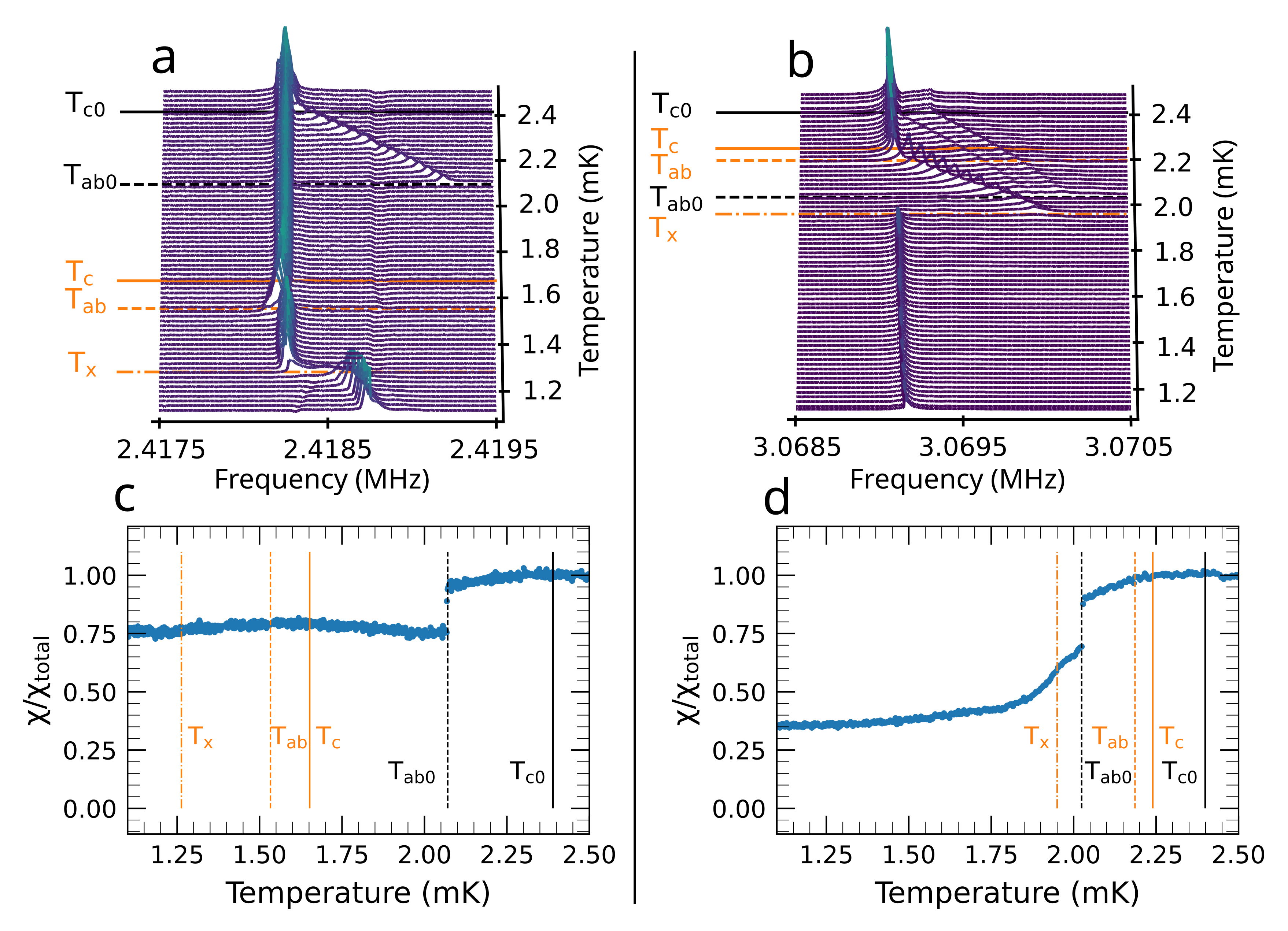}
\caption{\label{fig:4panel} Comparison of \3he{} NMR spectra and  total susceptibility between stretched and compressed aerogels. \3he{} data in $16\%$ stretched aerogel is shown in {\bf a} and {\bf c} taken on warming at 26.5 bar with small tipping angle ($\beta \approx 8^{\circ}$). Compressed aerogel is shown in {\bf b} and {\bf d} for $19\%$ compressed aerogel with small tipping angle ($\beta \approx 10^{\circ}$) with NMR spectra  taken on warming at 27 bar Ref. \cite{Zim.18}. In both cases the contribution from  pure \3he{} surrounding the aerogel is evident. Below the critical temperature $T_{c0}$ of the pure, bulk liquid, we observe the known positive frequency shift of the superfluid \Aph{}. At $T_{ab0}$ the pure component transitions to the \Bph{}. The frequency distribution of the pure spectrum is texturally broadened, and rapidly disappears into the noise floor. Consequently, in both samples at this temperature the total susceptibility decreases abruptly. Most importantly, the total susceptibility in the \Bph{} below $T_{ab}$ behaves very differently for the two phases, as is elaborated upon in the text.  The identification of the various superfluid phase transitions for \3he{} in aerogel are shown in orange.  They are based on frequency shift measurements, together with their tipping angle dependence, discussed in detail in Ref.\cite{Sco.23}.}
\end{figure*}   

In Fig.\,\ref{fig:4panel}, NMR spectra for small tipping angles from both the (\textbf{a, c}) $16\%$ stretched and (\textbf{b, d}) $19\%$ compressed aerogels are compared. The upper panels Fig.\,\ref{fig:4panel} \textbf{a}, \textbf{b} display the NMR spectra, while the lower panels \textbf{c}, \textbf{d} display the integrated total spectral weight of these spectra, normalized to just above the pure superfluid transition temperature $T_{c0}$. Each series of NMR spectra display five transition temperatures - two correspond to the transitions in the pure superfluid surrounding the aerogel sample. At $T_{c0}$, the pure liquid becomes the superfluid \Aph{} and is seen to have a positive frequency shift separating it from the spectral component of the liquid \3he{} in the aerogel, which is still in the normal state. At $T_{ab0}$ the pure liquid enters into the \Bph{} and the satellites disappear and contribute minimally to the overall spectral weight. The commensurate decrease in susceptibility between both samples at $T_{ab0}$ reflects the similar geometry of both sample arrangements.

The fraction of the superfluid imbibed in the aerogel constitutes the majority of the spectral weight of the NMR response. Below the aerogel superfluid transition temperature  $T_{c}$, the superfluid imbibed in either compressed or stretched aerogels is in the \Aph{}. In the stretched aerogel (\textbf{a}), the orbital angular momentum vector $\bm{\ell} \parallel \bm{\epsilon} \parallel \bm{H_{0}}$, producing a negative frequency shift comparable to what is expected in a slab confined system\cite{Aho.75a,Lev.13b,Li.14a}. In contrast, in the compressed aerogel (\textbf{b}) the orbital angular momentum is 2D-disordered with $\bm{\ell} \bot \bm{\epsilon} \parallel \bm{H_{0}}$, producing a small positive frequency shift\cite{Li.13}. 

Below the aerogel \textit{A} to \textit{B} transition temperature, denoted $T_{ab}$, both samples transition to the \Bph{}. This transition preserves the orientation of the orbital angular momentum relative to the axis of anisotropy. As a consequence in the stretched aerogel, with $\bm{\ell} \parallel \bm{\epsilon}$, the spectrum is effectively unshifted while in the compressed aerogel with $\bm{\ell} \bot \bm{\epsilon}$ there is a large positive frequency shift. 

At the orbital-flop temperature $T_{x}$ the orbital quantization axis $\bm{\ell}$ reorients by $90^\circ$ relative to the strain axis $\bm{\epsilon}$. As a consequence the different response of the NMR frequency shift to the structure of the aerogel is swapped between the compressed and stretched aerogel, with the compressed aerogel now effectively unshifted while the stretched aerogel experiences a strong positive frequency shift. Although the orbital flop transition has been previously identified in stretched aerogel samples in the \Aph{} without preplating\cite{Pol.12,Li.Thesis,Ngu.24}, it is noteworthy that in $16\%$ stretched aerogel it also occurs in the \Bph{} in an orientation-temperature order exactly reversed from the similar transition occurring in the \Bph{} in compressed aerogel.

The susceptibilities of the superfluid \Bph{} imbibed in the two aerogel samples are substantially different, as seen in the lower panels of Fig.\,\ref{fig:4panel} \textbf{c} and \textbf{d}. While the susceptibility of the compressed aerogel \Bph{} displays a gradual decrease with temperature commensurate with the growth of the suppressed order parameter\cite{Sha.01,Min.01}, the susceptibility of the \3he{} in the stretched aerogel remains effectively temperature independent to the lowest temperature, despite the growth of the of the zero-spin projection order parameter amplitude. As a consequence the enhanced susceptibility of planar aerogel is incompatible with both gap-distortion effects and isotropic impurity models\cite{Sco.23}. This is also one of the key results of the present work.

\section{Aerogel Structure Simulations}

\begin{figure*}
\centering
\includegraphics[width=120mm]{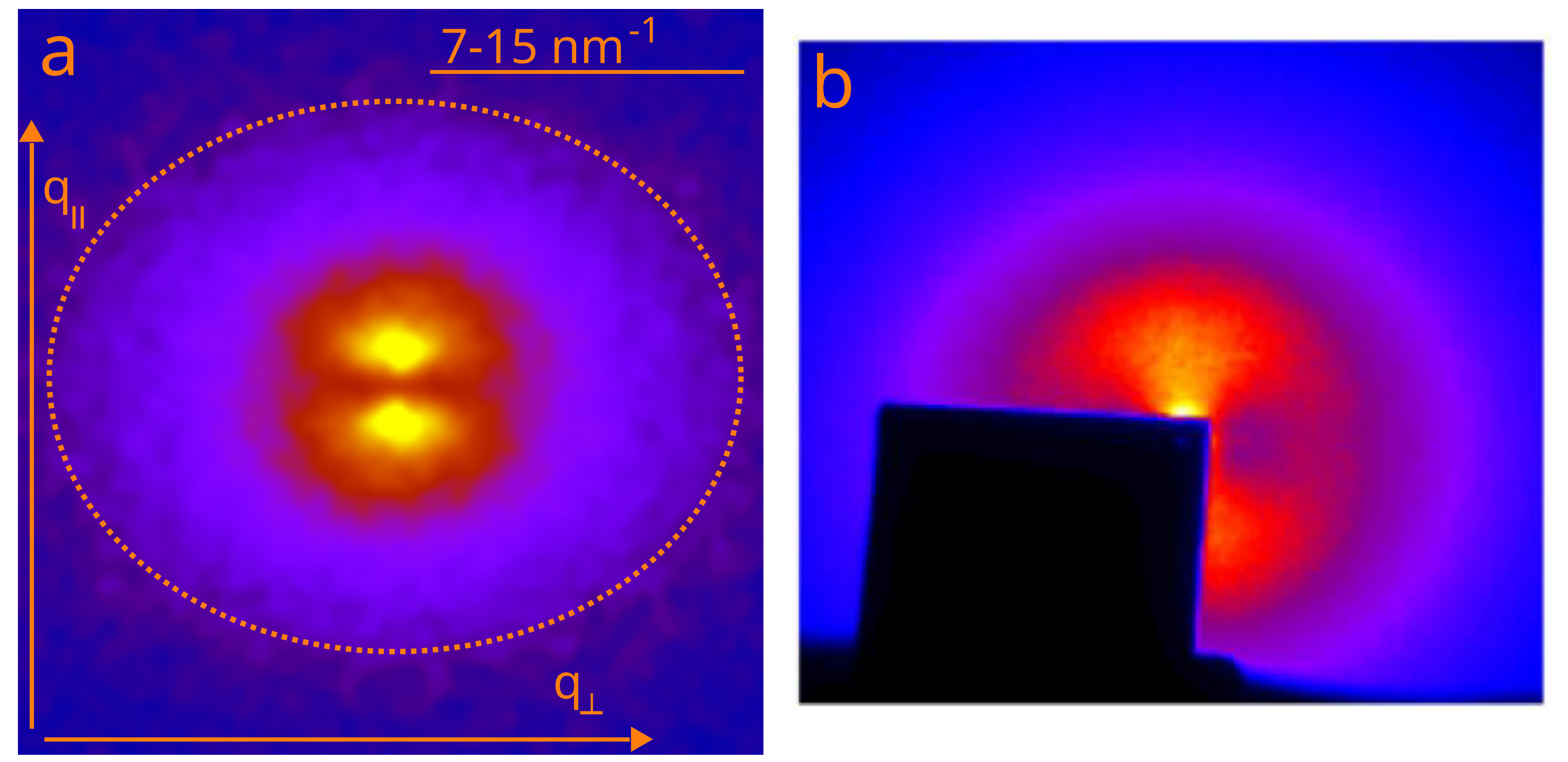}
\caption{\label{fig:saxfigure} Comparison of structure factors $S(\bm{q})$ in simulated anisotropic aerogel \textbf{a} with experimental measurement \textbf{b} via small-angle x-ray scattering.  The simulated structure factor of a stretched aerogel, \textbf{a}, \cite{Ngu.24} concurs with the SAXS data, \textbf{b} \cite{Pol.08}. The distinctive double-lobed 'dipole' pattern in the inner region appears in both, showing an excess of correlation at long length scales in the direction of the axis of anisotropy. However, there is also an outer ellipsoidal shape which indicates an excess of correlation perpendicular to the axis of anisotropy at short length scales. This shape is traced with an orange dashed line, and occurs on the order of $q = 7 - 15 \, nm^{-1}$, for a typical silica particle size of $1 - 2 \, nm$. The large black rectangle in \textbf{b} is a beamstop blocking the transmitted portion of the beam.} 
\end{figure*} 

Simulations of anisotropic silica aerogels performed with an anisotropically biased diffusion-limited cluster aggregation algorithm show that plane-like structures develop with unidirectionally biased diffusion\cite{Ngu.24}. The structure factor of such aerogels $S(\bm{q})$, for wavevectors $\bm{q}$, produces an image with similar features as the experimental x-ray scattering cross-section for a stretched aerogel. In these images, more intense scattering at small $\bm{q_{\parallel}}$  in the direction of the axis of anisotropy $\bm{\epsilon}\parallel \bm{q_{\parallel}}$, oriented towards the top of the page, appears as a double-lobed dipole pattern near the center of the image. The structure factor also shows an ellipsoidal pattern, rotated $90\deg$ from the dipole, extending further out in $\bm{q_{\bot}}$. There is a crossover between the dipolar and ellipsoidal regime at an intermediate length of roughly $50\,\mathrm{nm}$. The comparison between the simulated structure factor and the small-angle x-ray scattering validates the simulation from which the structure factor was calculated. Real-space images of the simulated stretched aerogels show the planar-like structures, while compressed aerogels have a nematic-like structure, for details see Ref. \cite{Ngu.24}. The latter have structures that orient the orbital component of the superfluid order parameter exactly opposite to that of stretched case. This opposite orientation is displayed in the data in Fig.\,\ref{fig:4panel}, which compares the frequency shift and susceptibility of $16\%$ stretched aerogel\cite{Sco.23} to $19\%$ compressed aerogel \cite{Zim.18} at comparable pressures and fields. 

\section{Structure and Transitions}

\begin{figure*}
\centering
\includegraphics{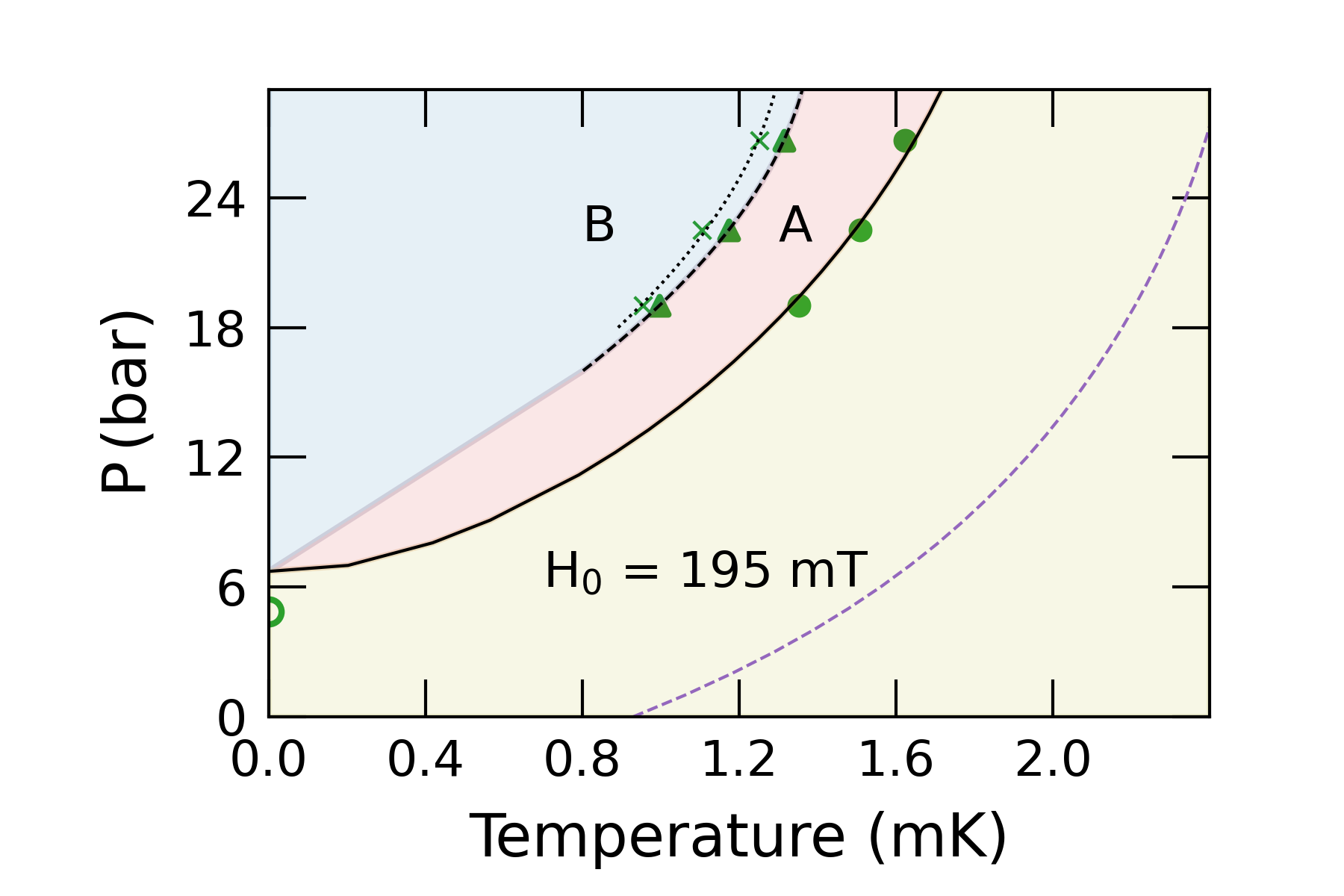}
\caption{\label{fig:pressphasediagram} The pressure phase diagram of $16\%$ stretched aerogel. The green circles denote the critical temperature $T_{c}$ in the aerogel with the solid black line as a fit to the critical temperatures\cite{Sau.03}; the green triangles indicate the $B$ to $A$ transition on warming in a field of approximately $195\,\mathrm{mT}$; the green $\mathrm{x}$ marks the temperature of the orbital flop transition \cite{Zim.18}. The purple dashed line is the pure superfluid critical temperature $T_{c}$, while the dashed black line denotes a guide to the eye for the $B$ to $A$ transition. The region of stable \Bph{} is in stark contrast to the measurements in a similar aerogel sample reported by Pollanen \etal{} \cite{Pol.12}}
\end{figure*} 

Calculations of the properties of \3he{} confined to a planar slab show a strong enhancement of the local magnetic susceptibility due to surface Andreev bound states\cite{Chu.09,Nag.09,Miz.12b}. In the highest degrees of confinement, it is calculated that the susceptibility averaged over the slab can recover the magnetic susceptibility of the normal liquid\cite{Nag.09}, although at such confinements the \Bph{} is no longer stable\cite{Zhe.17}. In the simulated aerogels, the planar structures are similarly closely spaced with a peak in the planar density modulations on the order of $50-100\,\mathrm{nm}$. This length is comparable to the size of the slab necessary to fully recover the normal state susceptibility. As a consequence, the likely cause of the near-constant susceptibility in stretched aerogel is similar bound states formed by planar structures. In contrast to both slab confinement and prior experiments performed in comparably stretched aerogel samples\cite{Pol.12,Dmi.20a}, the superfluid displays a large region of stable \Bph{}, Fig.\,\ref{fig:pressphasediagram}. The porosity of the aerogel structure versus the non-porous slabs likewise may play a role in the stability of the \Bph{} at such confinement.

\begin{figure*}
\centering
\includegraphics{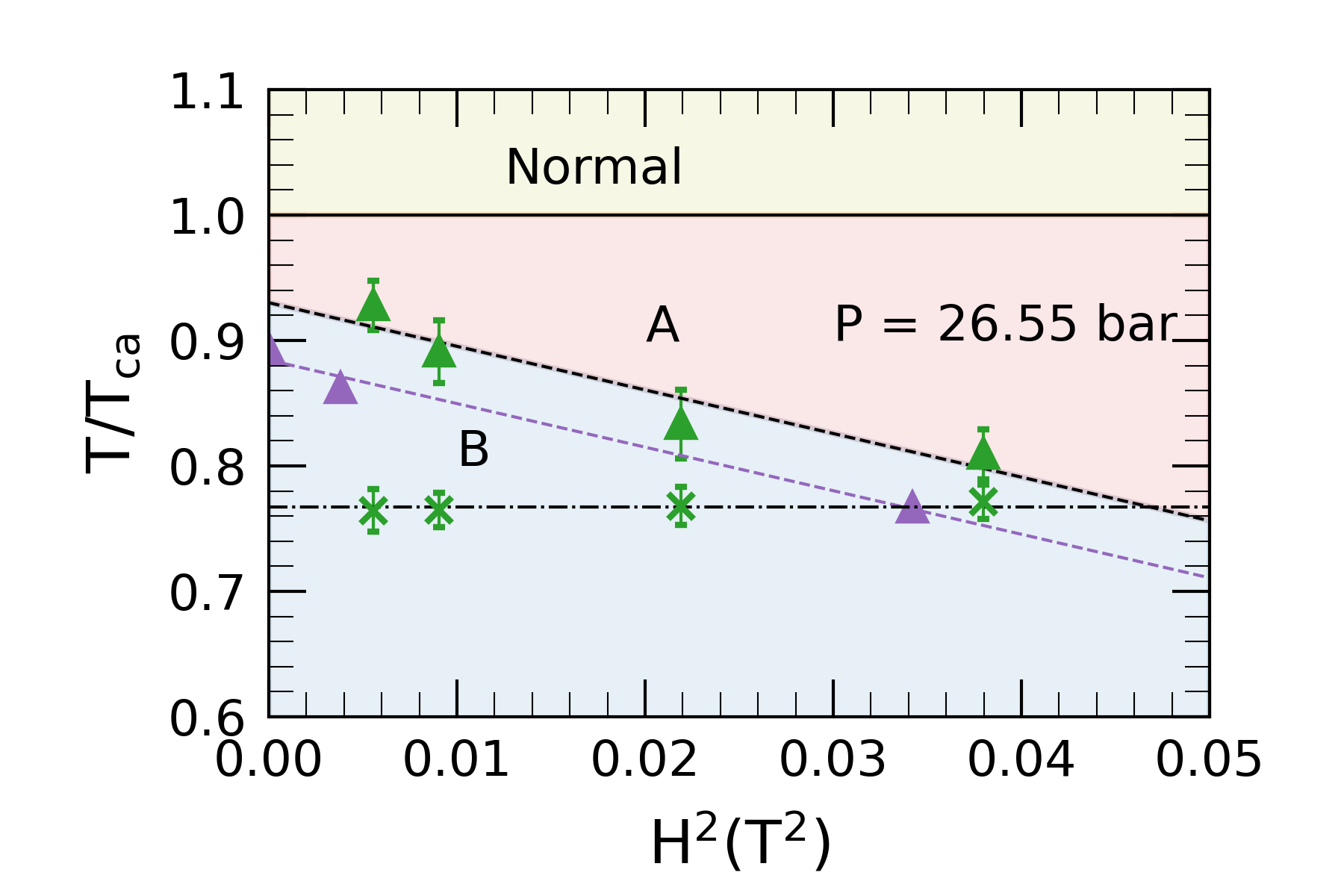}
\caption{\label{fig:magphasediagram} The magnetic phase diagram taken at $26.6\,\mathrm{bar}$ for \3he{} in $16\%$ stretched aerogel. The green triangular trace indicates the $B$ to $A$ transition in the superfluid on warming with the black dashed line a quadratic magnetic field fit to these transitions; the green $\mathrm{x}$ markers denote the field-independent orbital flop transition. The purple trace is the $B$ to $A$ transition on warming in the pure superfluid with the purple dashed line a quadratic fit to the data\cite{Sch.Thesis, Gre.86}. The comparable slopes stand in direct contrast to the temperature independence of the susceptibility equal in both  $A$ and $B$ phases.}
\end{figure*}

The crossover in the structure seen in the simulations has been identified as being directly responsible for the orbital-flop, a reorientation of the angular momentum axis $\bm{\ell}$, of the superfluid order parameter\cite{Ngu.24}, as studied in compressed aerogel\cite{Zim.18}. This transition is driven purely by the structure and, as expected for a transition in the orbital degrees of freedom, does not couple to the magnetic field. This is seen in Fig.\,\ref{fig:magphasediagram} and has also been reported in compressed aerogel. Consistently, no change in the magnetic susceptibility would be expected across this magnetic field-independent orbital-flop transition.

In contrast, varying the magnetic field applied to the sample shows suppression of the $B$ to $A$ transition temperature, $T_{ab}$, with increasing field as shown in Fig.\,\ref{fig:magphasediagram}. For the typically first-order $B$ to $A$ transition, the Clausius-Clapeyron relation requires
\begin{equation}
\frac{d T_{ab}}{d H^{2}} = - \frac{1}{2}\frac{\chi_{A} - \chi_{B}}{s_{A} - s_{B}}.
\end{equation}For approximately constant susceptibility in a $16\%$ stretched aerogel this transition must be close to isentropic, approximating a second order transition. Similar to the enhanced susceptibility discussed above, this is  analogous to the behavior expected in strongly planar-confined \3he{}. Under this confinement the transition is predicted to be either second-order or weakly first-order, dependent on confinement, magnetic field, and material parameters\cite{Har.88,Vor.07,Miz.12b}. Imbibing \3he{} in aerogel alters the effective material parameters of the superfluid\cite{Thu.98,Bar.02,Sau.03,Aoy.07}, potentially further approaching a second-order $AB$ transition. Additionally, it is noteworthy that the slope of this transition temperature dependence  on the square of the magnetic field is very similar to pure \3he{}, shown in Fig.\,\ref{fig:magphasediagram}.

\section{Summary}

Computational studies in conjunction with x-ray scattering experiments have revealed information about the microstructure of anisotropic aerogel. In particular, aerogel with stretched anisotropy has been shown to contain large plane-like structures. The thermodynamics of the \textit{A} to \textit{B} transition of superfluid \3he{} imbibed in such a planar aerogel are qualitatively similar to theoretical expectations for the superfluid transition when imbibed in a thin planar slab. This is manifest both as an enhancement to the magnetic susceptibility and a near-isentropic \textit{A} to \textit{B} transition. The structure of the aerogel exhibits a crossover in its preferred direction at intermediate lengths, driving a reorientation transition in the orbital angular momentum of the superfluid when this crossover length is commensurate with the superfluid coherence length. The character of these transitions is a consequence of the anisotropic structure of the aerogel, underscoring the novel physics present in impure unconventional superfluids. We would like to thank T. Mizushima, J.A. Sauls, and A.B. Vorontsov for helpful discussions. This work was supported by NSF DMR-2210112.

\bibliography{Scott_References}

\end{document}